\documentclass[aps,prb,reprint,twocolumn,superscriptaddress,showpacs]{revtex4-1}
\usepackage{amsmath,amssymb,amsfonts,amsthm}
\usepackage{graphicx}
\usepackage{bbm}
\usepackage{dcolumn}   
\usepackage{epstopdf}
\usepackage{subfig}

\begin{document}

 \newcommand{\breite}{1.0} 

\newtheorem{prop}{Proposition}
\newtheorem{cor}{Corollary}

\newcommand{\be}{\begin{equation}}
\newcommand{\ee}{\end{equation}}

\newcommand{\bea}{\begin{eqnarray}}
\newcommand{\eea}{\end{eqnarray}}

\newcommand{\Reals}{\mathbb{R}}     
\newcommand{\Com}{\mathbb{C}}       
\newcommand{\Nat}{\mathbb{N}}       

\newcommand{\id}{\mathbbm{1}}    

\newcommand{\Real}{\mathop{\mathrm{Re}}}
\newcommand{\Imag}{\mathop{\mathrm{Im}}}

\def\O{\mbox{$\mathcal{O}$}}   
\def\F{\mathcal{F}}			
\def\sgn{\text{sgn}}

\newcommand{\deo}{\ensuremath{\Delta_0}}
\newcommand{\dea}{\ensuremath{\Delta}}
\newcommand{\ak}{\ensuremath{a_k}}
\newcommand{\ad}{\ensuremath{a^{\dagger}_{-k}}}
\newcommand{\sx}{\ensuremath{\sigma_x}}
\newcommand{\sz}{\ensuremath{\sigma_z}}
\newcommand{\spl}{\ensuremath{\sigma_{+}}}
\newcommand{\smi}{\ensuremath{\sigma_{-}}}
\newcommand{\alk}{\ensuremath{\alpha_{k}}}
\newcommand{\bk}{\ensuremath{\beta_{k}}}
\newcommand{\ok}{\ensuremath{\omega_{k}}}
\newcommand{\vd}{\ensuremath{V^{\dagger}_1}}
\newcommand{\vi}{\ensuremath{V_1}}
\newcommand{\vo}{\ensuremath{V_o}}
\newcommand{\zc}{\ensuremath{\frac{E_z}{E}}}
\newcommand{\xc}{\ensuremath{\frac{\Delta}{E}}}
\newcommand{\xd}{\ensuremath{X^{\dagger}}}
\newcommand{\aok}{\ensuremath{\frac{\alk}{\ok}}}
\newcommand{\tpw}{\ensuremath{e^{i \ok s }}}
\newcommand{\tpe}{\ensuremath{e^{2iE s }}}
\newcommand{\tmw}{\ensuremath{e^{-i \ok s }}}
\newcommand{\tme}{\ensuremath{e^{-2iE s }}}
\newcommand{\epls}{\ensuremath{e^{F(s)}}}
\newcommand{\emis}{\ensuremath{e^{-F(s)}}}
\newcommand{\epl}{\ensuremath{e^{F(0)}}}
\newcommand{\emi}{\ensuremath{e^{F(0)}}}

\newcommand{\lr}[1]{\left( #1 \right)}
\newcommand{\lrs}[1]{\left( #1 \right)^2}
\newcommand{\lrb}[1]{\left< #1\right>}
\newcommand{\nbt}{\ensuremath{\lr{ \lr{n_k + 1} \tmw + n_k \tpw  }}}

\newcommand{\om}{\ensuremath{\omega}}
\newcommand{\dw}{\ensuremath{\Delta_0}}
\newcommand{\wbp}{\ensuremath{\omega_0}}
\newcommand{\dv}{\ensuremath{\Delta_0}}
\newcommand{\vbp}{\ensuremath{\nu_0}}
\newcommand{\vplus}{\ensuremath{\nu_{+}}}
\newcommand{\vminus}{\ensuremath{\nu_{-}}}
\newcommand{\wplus}{\ensuremath{\omega_{+}}}
\newcommand{\wminus}{\ensuremath{\omega_{-}}}
\newcommand{\uv}[1]{\ensuremath{\mathbf{\hat{#1}}}} 
\newcommand{\abs}[1]{\left| #1 \right|} 
\newcommand{\avg}[1]{\left< #1 \right>} 
\let\underdot=\d 
\renewcommand{\d}[2]{\frac{d #1}{d #2}} 
\newcommand{\dd}[2]{\frac{d^2 #1}{d #2^2}} 
\newcommand{\pd}[2]{\frac{\partial #1}{\partial #2}} 
\newcommand{\pdd}[2]{\frac{\partial^2 #1}{\partial #2^2}} 
\newcommand{\pdc}[3]{\left( \frac{\partial #1}{\partial #2}
 \right)_{#3}} 
\newcommand{\ket}[1]{\left| #1 \right>} 
\newcommand{\bra}[1]{\left< #1 \right|} 
\newcommand{\braket}[2]{\left< #1 \vphantom{#2} \right|
 \left. #2 \vphantom{#1} \right>} 
\newcommand{\matrixel}[3]{\left< #1 \vphantom{#2#3} \right|
 #2 \left| #3 \vphantom{#1#2} \right>} 
\newcommand{\grad}[1]{\gv{\nabla} #1} 
\let\divsymb=\div 
\renewcommand{\div}[1]{\gv{\nabla} \cdot #1} 
\newcommand{\curl}[1]{\gv{\nabla} \times #1} 
\let\baraccent=\= 

\title{Polaronic model of Two Level Systems in amorphous solids}

\author{Kartiek Agarwal}
\affiliation{Physics Department, Harvard University, Cambridge, Massachusetts 02138, USA}
\email[]{agarwal@physics.harvard.edu}
\author{Ivar Martin}
\affiliation{Theoretical Division, Los Alamos National Laboratory, Los Alamos, New Mexico 87545, USA}
\author{Mikhail D. Lukin}
\affiliation{Physics Department, Harvard University, Cambridge, Massachusetts 02138, USA}
\author{Eugene Demler}
\affiliation{Physics Department, Harvard University, Cambridge, Massachusetts 02138, USA}


\date{\today}
\begin{abstract}

While two levels systems (TLSs) are ubiqitous in solid state systems, microscopic understanding of their nature remains an outstanding problem. Conflicting phenomenological models are used to describe TLSs in seemingly similar materials when probed with different experimental techniques. Specifically, bulk measurements in amorphous solids have been interpreted using the model of a tunneling atom or group of atoms, whereas TLSs observed in the insulating barriers of Josephson junction qubits have been understood in terms of tunneling of individual electrons. Motivated by recent experiments studying TLSs in Josephson junctions, especially the effects of elastic strain on TLS properties, we analyze interaction of the electronic TLS with phonons. We demonstrate that strong polaronic effects lead to dramatic changes in TLS properties. Our model gives a quantitative understanding of the TLS relaxation and dephasing as probed in Josephson junction qubits, while providing an alternative interpretation of bulk experiments. We demonstrate that a model of polaron dressed electronic TLS leads to estimates for the density and distribution of parameters of TLSs consistent with bulk experiments in amorphous solids. This model explains such surprising observations of recent experiments as the existence of minima in the energy of some TLSs as a function of strain and makes concrete predictions for the character of TLS dephasing near such minima. We argue that better understanding of the microscopic nature of TLSs can be used to improve properties of quantum devices, from an enhancement of relaxation time of TLSs, to creating new types of strongly interacting optomechanical systems.

 \end{abstract}

\maketitle

\section{Introduction}
\label{sec:Intro}
%

At low temperatures many physical properties of amorphous solids are dominated by an ensemble of Two-Level- Systems (TLSs) \cite{Phillips}. Measurements of the thermal, elastic, and dielectric properties of these materials \cite{zeller} have been successfully interpreted within the phenomenological model of atoms (or groups of atoms) tunneling in random double well potentials \cite{Anderson,Philold}. While these experiments did not provide a direct evidence for the nature of the tunneling objects, the small energies of TLSs were taken as corroboration of tunneling atoms rather than electrons. To contribute appreciably to thermal properties at a temperature of hundreds of mK, a significant number of TLSs should have energies of the order of GHz, which was considered inconsistent with the typical eV energy scale for electrons in solids (see e.g. WKB estimate of tunneling matrix elements \cite{Anderson}). However subsequent experiments on the effect of TLSs provided evidence that tunneling objects carry electric charge and spin, thus casting doubt on the picture of tunneling atoms. 

It is now known that TLSs in amorphous solids are very sensitive to even weak electric and magnetic fields \cite{Hunklinger,Enss,Fulde,Wurger}. Moreover, TLSs in the insulating layers of Josephson Junctions (JJ) were found to be rather strongly coupled to JJ qubits. As the energy of the JJ qubit is tuned, it comes into resonance with TLSs, leading to anti-crossing gaps of up to 30 MHz \cite{Ustinov,Ustinovold}. This strong coupling corresponds well with the interaction between typical electric fields inside JJs and a TLS having a dipole moment corresponding to an electron tunneling through a distance of a few Angstroms \cite{Martinis05,Yus}.The aim of this paper is to resolve the seeming discrepancy between the earlier bulk measurements, suggesting tunneling of heavy particles, and subsequent experiments, indicating the tunneling of electrons \cite{Andreev,Ioffe,sarma}. We show that the strong interaction between tunneling electrons and phonons leads to significant dressing of electrons. We demonstrate that this polaronic picture of TLSs not only describes both qualitatively and quantitatively the properties of TLSs as observed in JJ Qubits, but also provides a natural explanation for many parameters built into the phenomenological model of TLSs given by Anderson {\em et al.} \cite{Anderson}

Our theoretical model is motivated by recent experiments with TLSs in JJ qubits by Grabovskij {\em et al.} \cite{Ustscience}, which observed dramatic changes in the TLS energy under the application of mechanical strain to the sample. The dimensionless parameter characterizing this change $K =d log 2E/d\epsilon$, where $E$ is the TLS half-energy and $\epsilon$ is the strain (which in the simplest case can be understood as $\epsilon = \delta a/ a$, where $\delta a$ is the change of the lattice constant $a$), was found to be consistently in the range of $10^5$. Intriguingly, this variation of energy with applied strain is in complete accordance with the model of TLS-phonon interaction used to describe bulk experiments 
\begin{align}
H_{TLS} = E_z \sigma_{z} + \Delta \sigma_{x} + \gamma_{z} \epsilon \sigma_{z}.
\label{eq:1}
\end{align}
$\epsilon$ here is the strain at the position of the TLS. For TLS energies accessible via JJ devices ($\sim 10$ GHz), $K \sim 10^5$ implies $\gamma_z \sim$ 1 eV, which is precisely the value of $\gamma_z$ one deduces from, e.g., bulk measurements of thermal conductivity. Furthermore, in the range of strain applied, even a complete hyperbolic variation of the half energy $E= \sqrt{E^2_z + \Delta^2}$ through its minimum ($E_z = 0$) has been observed for some TLSs.  

While such large TLS-phonon coupling motivates the analysis of polaronic effects, one would expect that it naturally gives rise to a large rate of dephasing. This stands at odds with another important finding \cite{Ustinovold,unpub} which is that for most, if not all TLSs investigated at energies of $\sim 10$ GHz, the decoherence time $T_2$ is of the order of the relaxation time $T_1$, reaching nearly twice of $T_1$ for a few TLSs. As we will show, the bulk of such TLS-phonon coupling only leads to polaronic dressing without resulting in any anomalously large dephasing. 

The paper is organized as follows. To begin with, in Section \ref{sec:Micro}, we study the problem of a tunneling system in which the principle object is an electron. Typical barrier height ($\sim$ eV) and tunneling length ($\sim$ \AA) result in TLS energies at the scale of 1eV, clearly incompatible with observed TLS energies ($\sim$ GHz). We show that interaction of this confined electron with optical phonon modes strongly alters TLS asymmetry $E_z$ and suppresses $\Delta$ to result in energies in the GHz regime. We then explain how coupling to acoustic phonons can result due to the modification of the single-electron potential, as well as polaronic effects, whereby non-linearity in the optical phonon modes provide for a resultant interaction of the TLS with acoustic phonons. Our model allows us to show why $\gamma_z \sim 1$ eV, explain why $\gamma_x$ (coupling of strain $\epsilon$ to $\sigma_x$ neglected in Eq. (\ref{eq:1})) is insignificantly small, and derive a transparent estimate of the spatial density of TLSs and the distribution of parameters $E_z$ and $\Delta$. 

Because of strong renormalization due to electron-phonon coupling, TLS dynamics must be derived fully incorporating its polaronic nature. In Section \ref{sec:Dec}, we treat the strong TLS-phonon interaction via a non-perturbative variational approach (the partial polaron transformation \cite{Silbey}), that results in an effective Hamiltonian of the TLS that is only weakly correlated with the bath of phonons. This then allows us to treat the dynamics of the TLS in the phonon bath under the Born-Markov approximation, and analyze the effect of phonons on relaxation and dephasing of the TLS. We find, as mentioned, that pure dephasing due to phonons is negligible, and that the dominant decay process is via single phonon emission leading to an effective Fermi golden rule like decay rate, which is in quantitative agreement with experiments. 

From an experimental point of view, an intriguing signature of phonon induced decay is that a broad maximum (with possible variations due to TLS-TLS interactions) of relaxation time $T_1$ is achieved as TLS energy is tuned to a minimum by applying strain. This goes against the expectation that a fully symmetric double well (achieved at minimum $E$) should be more susceptible to decay from a noise that breaks this symmetry ($\gamma_z \epsilon \sigma_z$), but as we show, this is really a characteristic of the superohmic noise spectrum of strain noise \cite{Leggett}. Phonon induced decay can also be effectively controlled. In Section \ref{sec:crystal}, we present an experimental scheme based on band structure calculations that show how sizable phononic band gaps can be engineered in Al based superconducting qubits by appropriate modifications of device geometry, and that could result in significantly enhanced lifetimes of TLSs. 

In Section \ref{sec:intdec}, we discuss the effect of TLS-TLS interactions on the relaxation and dephasing properties of TLSs. We find that other TLSs contribute insignificantly to TLS relaxation, but can be expected to dephase the TLS in such a way that the decay of the off-diagonal TLS density matrix element is gaussian in time rather than simply exponential. This altered form of decay is a consequence of pure dephasing due to thermally activated TLSs that are typically slower than the TLS under examination (in the event that the TLS energy is a few times greater than the temperature of the system, which is usually the case in experiments). Furthermore, at temperature $T \sim 100$ mK, the time scale of this dephasing process is of the same order as the decay time due to phonons of TLSs with energy $2E \sim 10$ GHz. This explains why most observed TLSs at such energies tend to have $T_2$ of the same order as $T_1$. 

Finally, we point out that an automatic result of the polaronic suppression of tunneling, is the suppression of all TLS coupling to the environment through the tunneling ($\sigma_x$) operator. As a result, all TLS coupling to the environment is primarily through its dipole or $\sigma_z$ operator, which in the TLS eigenbasis, transforms to $(E_z/E) \sigma_z + (\Delta/E) \sigma_x$ This implies that the pure dephasing rate itself can be tuned proportionally to $E_z/E$, turning to zero at the TLS symmetry point. At this point, TLS decoherence is purely driven by relaxation ($T_2 \sim 2 T_1$, and as experiments find \cite{unpub}), and our phononic crystal scheme becomes extremely viable in significantly enhancing TLS coherence properties.

\section{Microscopic Model}
\label{sec:Micro}

\begin{figure}[ht]

\begin{center}
\includegraphics[width = 3 in]{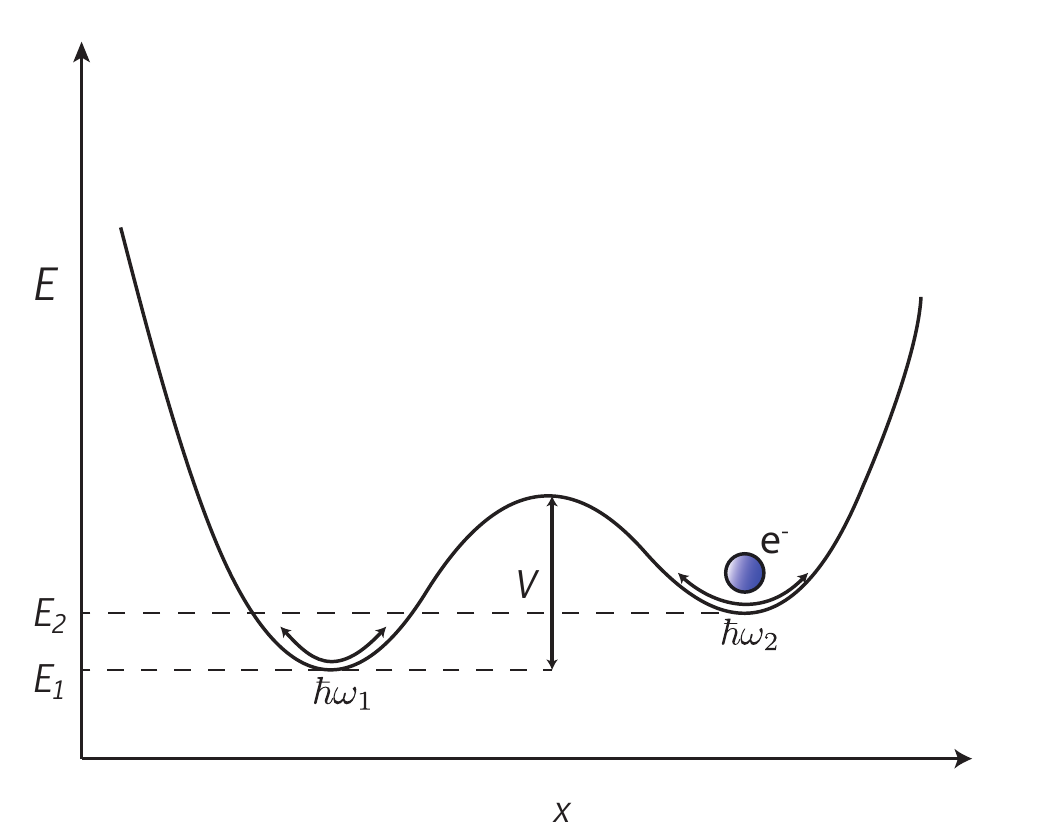}
\end{center}

\caption{An electron in a double well potential}
\label{fig:1}
\end{figure}

A valence electron typically finds itself sitting in a potential well structure provided by the electrostatic interaction with the nearest atoms. Given a disordered spatial configuration of such atoms, it may happen that more than one such potential well becomes accessible to an electron. In other words, if two such potential wells happen to be near enough (say a few $\AA$ apart), an electron can coherently tunnel between such potential minima, and it becomes important to understand the properties of the electron in such a double well potential. The double well itself can be characterized by the depths of the wells $E_1$ and $E_2$,  energies of zero point motion $\hbar \omega_1$ and $\hbar \omega_2$ in the respective wells, and an interwell separation $\Delta x$. Given that these structures are created by the atomic scale Coulomb interactions, we expect the scale of the energies involved to be on the order of an eV. This is in contrast to the energy scales used to describe the tunneling of atoms, $10^{-2}$ eV \cite{Philold}, that is of an order below the Debye frequency. For a sufficiently well aligned double well potential, one can work in the tight binding approximation of the two lowest lying energy eigenstates in either well, neglecting all higher energy levels and arrive at the following two level Hamiltonian for the electron in the double well structure,
\be
 H_o = \;
 \begin{pmatrix}
  E_1 + \hbar \omega_1 & \hbar \omega_0 e^{-\Gamma}  \\
  \hbar \omega_0 e^{-\Gamma} & E_2 + \hbar \omega_2 
 \end{pmatrix} . 
\ee
Here $\omega_0$ is of the order of $\omega_1, \omega_2$ and $\Gamma =  \sqrt{2 m V / \hbar^2} \Delta x $, $m$ being the mass of the tunneling object, and $V$ is the barrier height \cite{Phillips}.  For an electron, with $V \sim$ few eV, and $\Delta x$ of a few $\AA$, $\Gamma\sim1$.  In terms of the asymmetry $E_z^0 = E_1 + \hbar \om_1 - E_2 - \hbar \om_2$ and the tunneling strength $\Delta_0 = \hbar \om_0 e^{-\Gamma}$, one can effectively describe the system by the two level system (TLS) Hamiltonian, 
\be
H_o = E_z^0 \sz  + \Delta_0 \sigma_x .
\ee
Here $E_z^0$ and $\Delta_0$ are both of the order of an eV. The interaction of the qubit with the TLS is believed to be primarily an electric field-dipole coupling that can be described by  $S \sigma_z \tau_x$, where $\sigma$ and $\tau$ are the Pauli matrices corresponding to the TLS and the qubit respectively. Here, $S = e E_q (\Delta x) \cos \phi$ is the interaction strength that just represents the coupling of an electric dipole of charge $e$, length $\Delta x$, and orientation angle $\phi$ (the TLS), with the qubits electric field whose most significant matrix element is $E_q$, that between the ground and excited states (since the electric field of the qubit is proportional to the time-derivative of the phase difference across the Josephson Junction, it is very small in both eigenstates of the phase qubit, which are approximately the harmonic oscillator eigenstates in the phase variable). Such estimate of interaction was made in \cite{Martinis05} and is in reasonable agreement with the observed gaps in the qubit spectrum due to TLSs. Hence, the Hamiltonian that describes the combined system of qubit and TLS is 
\be
H_o + H_{TLS-Q} = E_z^0 \sz + \Delta_0 \sigma_x + S \sigma_z \tau_x.
\label{eq:q}
\ee

\subsection{Interaction with Longitudinal Optical Phonons}
\label{sec:LO}

Before we perform the analysis of dressing of electrons by high frequency phonons more rigorously using a variational approach in Section \ref{sec:Dec}, to gain some intuition we consider a simplified model of the interaction of the TLS with local Longitudinal Optical (LO) phonons whose equilibrium position is strongly affected by the state of the TLS. One can write down the Hamiltonian as follows,
\be
H_o + H_{op} = E_z^0 \sz + \Delta_0 \sigma_x + \frac{1}{2} k x^2 + \frac{p^2}{2m} +  \lambda x (\sz + c) , 
\label{eq:op} 
\ee
where one can think of $E_z^0$ as coming from the potential created by atoms frozen into an equilibrium position $x=0$ and $H_{op}$ describes the energetics of the local phonon due to interaction with other atoms in the material and the electron in the double well. 
%

Since we expect the energy of the renormalized TLS to be in the region of $10$ GHz (and as the renormalization procedure will yield) which is much smaller than $\om_D$, the frequency of the phonons, we expect the phonon, to always remain in its ground state, even as the TLS changes its internal state. This key assumption will be justified later in Section \ref{sec:Dec}. Integrating out the phonon then requires a minimization of the energy, which is achieved by the phonon instantaneously positioning itself according to which well the electron is occupying. Since the coupling of the phonon to the TLS is such that it has different energies for different electron positions (due to generally nonzero asymmetry constant $c$), the energy asymmetry of the resulting Hamiltonian is also modified by a factor proportional to the polaronic shift $E_p$. In particular, $ \avg{x}= - \lambda (\sz + c) /k $ and $E_p = \lambda^2 / k$. This polaronic shift has been estimated to be of the order of an eV \cite{polar} in the context of Si-MOSFETs, with the material - amorphous SiO$_2$. 

One must also remember that this procedure also involves a renormalization of the tunneling strength $\Delta_0$ which arises due to modification of $\Gamma$ by an additional factor of the overlap of the TLS-state-dependent phonon wavefunctions. For harmonic oscillator wavefunctions, this factor is simply $e^{-(x_1 - x_2)^2/4l^2} = e^{- E_p / \om_D}$ where $l$ is the zero-point motion length associated with an oscillator of frequency $\om_D$. With that, the effective low-energy Hamiltonian becomes 
\begin{align}
H_{TLS} &= \left( E_z^0 - c E_p  \right) \sz + \Delta_0 e^{- E_p/ \om_D} \sigma_x , \nonumber \\ 
E_p &= \lambda^2 / k . 
\label{eq:temp}
\end{align}

Note that even though $E_z^0$ and $E_p$ both are of the order of an eV, for some TLSs these can nearly cancel each other. Further, since $E_p / \om_D \sim 10$, $\Delta = e^{-10} \Delta_0$, and $\Delta_0$ can as well be of the order of $10^{-5} eV$. Thus, renormalization due to optical phonons can in principle result in TLSs with energies of the order of $10^{-5}$ eV, which is the energy scale at which JJ qubits operate and observe TLSs. Again, the idea here is that the renormalized energies of the TLS (determined by many different phonon modes collectively) are much smaller than the frequency of the phonon ($\om_D$) which means that the phonon always remains in its ground state. We want to alert the reader that our approach is in contrast with the usual idea of the Born Oppenheimer approximation where it is the electron that instantaneously assumes the ground state wavefunction.

\subsection{Density and Distribution of TLSs}

As seen above, due to the coupling to LO phonons, the asymmetry energy is shifted by $c E_p$. We expect $c$ to be a uniformly distributed number of order 1 for optical phonons. Therefore, we can obtain a simple estimate of the density of the TLSs which have an energy asymmetry of about $10^{-5}$ eV. The cancellation of two randomly distributed quantities $c E_p$ and $E_z^0$, both of which are of the order of 1 eV, to a value of about $10^{-5}$ eV happens with a probability of $10^{-5}$. 
The total energy of TLS is a combination of the energy asymmetry and the tunneling. However,  due to the polaronic renormalization of the tunneling matrix element (for the typical electron-phonon coupling strength in polar insulators, $E_p/\om_D \sim 10$), its contribution to total energy is expected to be smaller or comparable to the asymmetry contribution \footnote{It is important to recall that there is no prefactor $c$ in the renormalization of the tunneling strength $\Delta = \Delta_0 e^{- E_p / \om_D}$. Thus, we usually expect that a factor of $e^{- E_p /\om_D} \sim e^{-10}$ renders $\Delta$ small or comparable to $E_z$} and hence does not affect our estimate for the probability. Given that it is $E_z$ that primarily dictates the energy of the TLS, the purely probabilistic cancellation of $c E_p$ and $E^0_z$, implies that the energies of the TLSs are uniformly distributed in the range of energies $E_z < E_z^0$ yielding a constant density of states for the TLSs, which is essential for obtaining the the linear temperature dependence of the specific heat in bulk amorphous solids, and is in agreement with Anderson {\em et al.} \cite{Anderson} and Phillips \cite{Philold}. It is also important to emphasize here that such TLSs are expected to exist for a wide range of energies (up to eV), but only a small fraction of such TLSs are observable in JJ qubit experiments due to the limited range in which the qubit's energy can be tuned. 

What is novel in our approach however is that we can go a step further and give a quantitative prediction of the density of states of these TLSs. If we take such a double-well configurations to occur once in every region of about $ (10 \AA)^3$, this gives us a density of about $10^{-8}$ TLSs per $\AA^3$ for TLSs of energy of up to $10^{-5}$ eV. Using the fact the energy of the TLSs are uniformly distributed, density of states per unit volume of TLSs obtained is $10^{46} J^{-1} m^{-3}$ which is indeed in the range of   experimental estimates ($10^{45} - 10^{46} J^{-1} m^{-3}$) of density of TLSs in bulk amorphous solids \cite{Phillips}. 

\subsection{Coupling of TLS to acoustic phonons}
 \label{sec:AC}
 
 Acoustic phonons can couple to the $\sigma_z$ term of the TLS by modulating either $E_p$ or $E_z^0$ and both forms of coupling may be equally important. First, we consider the coupling to $E_z^0$.  We can expect that a strain field $\epsilon$ would modify the energy of either well by a factor of $E_o^z \epsilon$  which would give the required coupling strength to acoustic phonons of the form $\gamma_z \epsilon \sigma_z$, with $\gamma_z \sim E_o^z \sim 1$ eV as long as there is an asymmetry between the wells to begin with.
 
The effect of strain on $E_p$ is captured by the Gruneizen parameter $\eta = d \log \omega / d \epsilon \sim 1$ which measures the logarithmic change in energy of an individual phonon mode upon application of uniform strain. Since $E_p$ is determined by optical phonons, which effectively see the strain due to long wavelength acoustic phonons as uniform strain, the coupling to phonon fields can also be written as $E_p \eta \epsilon$ where, as usual, $\epsilon$ is the local strain field of acoustic phonons on the TLS.  A more detailed derivation of this result is given in Appendix \ref{sec:grun}.

As we said before, we require $c E_p$ and $E_z^0$ to cancel each other to give $E_z$ in the region of $10^{-5}$ eV ($\sim$ 10 GHz) to result in TLSs of energies that are accessible in JJ Qubit experiments. On the other hand, changes in both $c E_p$ and $E_z^0$ due to strain are of the order of 1, i.e, $\delta E_z^0 / E_z^0 \epsilon \sim 1$ and $\delta E_p / E_p \epsilon \sim 1$. Since the changes in the two quantities are generically different, the change in $E_z$, $\delta E_z = \delta E_z^0 - c \delta E_p$  is of the order of eV times strain, i.e. $\delta E_z \sim (eV) \epsilon$. We see that the dimensionless coupling parameter $K = \delta E / ( E \epsilon) \sim \delta E_z / E_z \epsilon$ is essentially a direct comparison of these two energy scales - eVs and 10 GHz, which so happens to be the same order of $10^5$ that is observed in experiments \cite{Ustscience}. We also expect that $E_z^0$ and $c E_p$ nearly cancel each other for some value of strain as it is slowly increased/decreased giving us a minimum value of $E = \sqrt{\Delta^2 + E^2_z}$ at some strain value, as is also seen in experiments \cite{Ustscience} (see Fig. \ref{fig:strain}). Finally, it is also easy to see that the change of $\Delta = \Delta_0 e^{- E_p / \om_D}$ due to strain is much less significant as $ d \log \Delta / d \epsilon \sim \eta E_p / \om_D \sim 10$, which is much less than $10^5$ figure that comes from the change in $E_z$ as a result of the change in $E_p$ or $E_z^0$. This explains the absence of a $\gamma_x \epsilon \sigma_z$ term in Eq. (\ref{eq:1}) 

Before concluding this discussion, we would like to contrast this analysis of the TLS-phonon coupling that arises for the electronic TLSs to the case of TLSs resulting from the motion of atoms or group of atoms. To find reasonable values of the tunneling strength, such as $\Delta \sim 10^{-5}$ eV, WKB estimates \cite{Philold} require that the atoms live in a potential landscape with valleys of depths of the order of $10^{-2}$ eV. For deeper valleys/barriers, $\Delta$ is quickly diminishes to irrelevantly small values. Since phonons interact with the TLS by modifying the shape of these valleys, one should expect the TLS-phonon interaction of the atomic TLS to be of the order of $10^{-2}$ eV times the strain. As we know, experiments clearly suggest a much larger value of this interaction, it is $\gamma_z \epsilon \sigma_z$, with $\gamma_z \sim$ eV. This discrepancy in the magnitudes of TLS-phonon interaction and the scale of the potential landscape that the constituent atoms occupy in the atomic model of TLSs should be a cause of concern as both scales correspond to the same physics. In our approach, this issue is naturally resolved as a confined electron is already associated with eV energy scales, and consequently $\gamma_z \sim$ eV is well justified. 


\section{Decoherence and Relaxation of the TLS by phonons}
\label{sec:Dec}

TLS decoherence  ($T_2$) and relaxation ($T_1$) times have been measured using the coherent transfer of state information between the TLS and the Qubit \cite{Ustinov}. Experiments find TLSs with decoherence times $T_2$ almost always of the order of $T_1$. Furthermore, as the TLS energy is tuned by applying strain, experiments find that at the point where the TLS energy is minimum, pure dephasing is negligible and $T_2 \approx 2 T_1$ \cite{unpub}. As mentioned before, it is necessary to take into account polaronic effects in order to properly discuss decoherence properties of TLSs. The crucial point is that given the large TLS-phonon coupling strengths, as found from experiments on both bulk amorphous solids and in JJ Qubit experiments, the phonons cannot be relegated to being a perturbative part of the environment but actually {\em dress} the electron significantly. It is this polaronic TLS with its residual interaction with phonons that in fact, is nearly immune to pure dephasing by phonons, allowing for the observation of TLSs with $T_2$ of the order of $T_1$. Furthermore, we will see that a natural result of the renormalization procedure is that the TLS can couple to the environment only through its dipole operator ($\sigma_z$ in the charge basis). In the TLS eigenbasis, this coupling is of the form -- $(E_z/E) \sz + (\Delta/E) \sigma_x$, and thus, as the TLS is tuned to its symmetry point $E_z = 0$, pure dephasing becomes negligible and one finds $T_2 \sim 2 T_1$. We will also show that a properly applied Fermi's Golden Rule using the strength of TLS-phonon coupling discussed earlier, can correctly capture the magnitude of the relaxation rate at low temperatures. An intriguing signature of phononic decay is that one should find an increase of $T_1$ as TLS energy is tuned to a minimum by virtue of the TLS-phonon coupling having a superohmic spectrum.

\subsection{Polaron Transformation}
\label{sec:pol}

To treat the problem of dressing of TLS by phonons more rigorously than we presented in Section \ref{sec:LO}, we use the partial polaron transformation approach developed by Silbey and Harris \cite{Silbey,Silbey2}. The advantage of this approach is that it does not presuppose any separation of time scales between phonons and the TLS, but allows for a systematic renormalization of TLS parameters and TLS-phonon coupling. The resultant Hamiltonian can then be used as an effective model to study the dynamical properties of the TLS (cf. \cite{dyna,cao} where such a procedure is shown to compare well with more rigorous methods). We will also see that this approach indeed consistent with the analysis presented in Section \ref{sec:LO}. To begin with, we consider the following Hamiltonian - which involves a generic coupling to phonons (acoustic or optical) - 

\be
H = E^{o}_{z} \sz + \Delta_0 \sx + \sum_k \; (\sz + c_k) \; g_k \; ( \ak + \ad ) + \sum_k \; \ok a^{\dagger}_k \ak . 
\label{eq:H}
\ee

This is the Hamiltonian of a TLS coupled to a bath of phonons. Note that $E_z^0$ and $\Delta_0$ are still the unrenormalized TLS parameters, in the eV range as described before. Also, hermiticity of the Hamiltonian requires $g_k = g^*_{-k}$ and $c_k = c^*_{-k}$. Guided by the discussion in the previous section, and the form of the coupling deduced from experiments, the form of the coupling parameter $g_k$ is given by the term -- $\gamma_z \epsilon(\vec{r}) \sigma_z$ where $\epsilon(\vec{r})$ is the local strain field, which contains all wavevectors up to $k_D$ in the Fourier space\footnote{One may have also included explicitly, as before, coupling to additional optical phonon modes with coupling $\lambda x \sigma_z$ as in Section \ref{sec:LO}, but, in fact, such optical phonon modes are already incorporated in the analysis of section \ref{sec:pol} where they simply arise as zone boundary ($k \rightarrow k_D$) acoustic phonons, with the coupling to phonon displacement $x$, $\lambda \sim \sqrt{2} \gamma_z k_D$}. Also, in principle, the strain field has a tensor structure, and correspondingly so does $g_k$. For the purposes of this discussion however, what is important is only the power of $k$ in the TLS-phonon coupling, and so we do not introduce the full tensorial structure here. For more discussion on this, see Appendix \ref{sec:TLSinteraction}. 

Now, $g_k (\ak + \ad)  = \gamma_z \epsilon_k$ giving $g_k = \gamma_z k \sqrt{\hbar/(2 MN \om_k)}$ where $\epsilon_k$ is the component of the strain field with wavevector $k$, and $MN$ is essentially the total mass of the atoms in the system. Also, $c_k$ is significant only for high frequency phonons near the Debye frequency (acoustic or optical) which have a wavelength comparable to the size of the TLS, that is acoustic phonons near the Debye frequency and optical phonons. We now transform the Hamiltonian using a generalized version of the polaron transformation, 
\begin{align}
& U = exp \left(  \sum_k \; ( \ak - \ad ) \frac{1}{\ok} ( \alk \sz + \bk ) \right) , \nonumber \\
& U^{\dagger} \; \ak \; U = \ak - \frac{\alk^*}{\ok} \sz - \frac{\bk^*}{\ok} , \nonumber \\
& U^{\dagger} \; \ad \; U = \ad - \frac{\alk^*}{\ok} \sz - \frac{\bk^*}{\ok} . 
\label{eq:trans}
\end{align}

As is clear from the form of Eq. (\ref{eq:trans}), the transformation is a matter of simply shifting the phonons according to which state the TLS occupies.  The Hamiltonian at this step is given in Eq. (\ref{eq:Ham1}). In what follows, we find the optimal values for $\alpha_k$ and $\beta_k$ by minimizing the Bogoluibov-Peirles upper bound on the free energy (refer to Appendix \ref{sec:polaron} for more details). The result of this is the effective Hamiltonian 
\begin{align}
\tilde H &= (E_z^0 - \sum_k 2 c_k \frac{g_k^2}{\ok} ) \sz + \dea \sx + \sum_k (g_k - \alk) (\ad + \ak) \sz \nonumber \\
&+ \deo ( e^{-2 \alpha} - \left< e^{- 2 \alpha} \right>_T) \spl + \deo ( e^{ 2 \alpha} - \left< e^{2 \alpha} \right>_T) \smi ,
\label{eq:final}
\end{align}
with the following pertinent equations for $\alpha_k$, $E_z$ and $\Delta$ 

\begin{align}
\alpha &=  \sum_k \alpha_k/\om_k (a_k - \ad) ,\nonumber \\
\alk &= g_k \Bigg/ \left(1+ \frac{2 E}{\ok} \left(  \lrs{\Delta/E} \tanh{\beta E} \coth{\beta \om_k/2} \right) \right) ,\nonumber \\
\Delta &= \Delta_0 \left< e^{2 \alpha} \right>_o = \Delta_0 \exp \left(-2 \sum_k \frac{\alk^2}{\ok^2} \coth{\left( \frac{\beta \ok}{2} \right)} \right), \nonumber \\
E_z &= E_z^0 - \sum_k 2 c_k g_k^2 / \om_k .
\label{eq:det}
\end{align}

Here, $E$ is the half of renormalized TLS energy, $E = \sqrt{E^2_z + \Delta^2}$, $\Delta$ is the ``mean field" renormalized tunneling strength, while the final two terms in Eq. (\ref{eq:final}) serve as perturbations. One can see that the factor renormalizing $E_z$ has the same form as in Eq. (\ref{eq:temp}), and so we suggestively denote factor renormalizing the asymmetry by $c E_p$ as before. An important result of these equations is that renormalization of $\Delta_0$ is primarily determined by high frequency phonons. To see this, one needs to note that for $\om_k \gg E$, $\alpha_k \sim g_k$ which goes as $\sqrt{k}$. The $\om^2_k$ in the denominator in the exponent of the expression for $\Delta$ in Eq. (\ref{eq:det}) is naturally cancelled by the density of states for phonons, which itself goes as $\om^2_k$. Thus, due to the form of $\alk$, we see that this exponent increases linearly in phonon frequency $\om_k$. As a consequence, the renormalization of $\Delta_0$ is primarily determined by high frequency phonons. One can then pull out a factor of $\om_k$ in Eq (\ref{eq:det}) for $\Delta$, replacing it with $\om_D$, and find that $\Delta \sim \Delta_0 e^{- E_p / \om_D} $ where $E_p = \sum_k 2 g^2_k/ \om_k$ (compare with Eq. (\ref{eq:temp})), which is the result we argued for earlier.

Another consequence of this fact is that since the thermal occupation of high frequency phonons is nearly unaffected by the change in temperature; even though $\alpha_k$ and $\Delta$ are really part of a self consistency equation, $\Delta$ essentially does not change with temperature. This non-dependence on temperature is obvious for $E_z$, and this result is consistent with the experimental findings\cite{Ustinov} which report no significant change of TLS energies with temperature. Further, one can evaluate the value of $\avg{e^{2 \alpha}}$, the factor that renormalizes $\Delta_0$ using the coupling strength to acoustic phonons as estimated from experiments, i.e., $\gamma_z \sim 0.5 eV$ (and parameters for Al such as density $\rho = 2700$ kg/$m^3$, Debye frequency $\om_D = 4 \times 10^{13}$ rad/s , and transverse wave velocity $v_t = 3000$  m/s) and find this factor to be of the order of $e^{-10} \sim 10^{-5}$, that we already used above. Thus, as mentioned earlier, one can arrive at similar estimates of the suppression of the tunneling strength $\Delta_0$ due to known estimates of the coupling to optical phonons (as in Section \ref{sec:LO}), or what is known of the TLS-phonon coupling from JJ qubit experiments and earlier bulk experiments. 

It is also important to note that this huge suppression in fact applies to all couplings that the TLS can have to the environment through its $\sigma_x$ operator. This can be understood by observing that any operator coupled to $\sigma_x$ in the Hamiltonian (\ref{eq:H}) (before polaron transformation) will acquire a factor $\avg{e^{-2\alpha}}$ after the polaron transformation as in eq (\ref{eq:final}). This then implies that the only relevant coupling of the TLS to its environment is through its dipole operator, $\sigma_z$. When we transform into the TLS eigenbasis, this coupling appears as $(E_z/E) \sz + (\Delta/E) \sigma_x$, and thus the coupling to the $\sz$ operator in the eigenbasis can be tuned as $(E_z/E)$. In particular, at the symmetry point, where $E_z = 0$, this coupling reduces to zero, thus making the TLS nearly immune to pure dephasing, which explains why experiments \cite{unpub} commonly find that the decoherence time $T_2$ reaches twice of $T_1$ at this symmetry point. 


Finally, it is important to note that the behavior of $\alk$ with phonon frequency $\om_k$ changes at the scale given by the typical frequency of fluctuations in the renormalized TLS's internal state, given by $\Delta \left( \Delta/E \right)$. For $\om_k \gg \Delta^2/E$, $\alk$ rapidly approaches $g_k$, whereas for $\om_k \ll \Delta^2/E$, $\alk$ approaches zero. Recall from Eq. (\ref{eq:trans}) that $\alk$ is associated with displacements of phonons of frequency $\om_k$ in accordance with the TLS's internal state. This behavior of $\alk$ with $\om_k$ thus corresponds to the fact that phonons whose frequencies are greater than the renormalized TLS tunneling rates, are able to (near) instataneously adjust themselves to the TLS's internal state (corresponding to the analysis presented in Section \ref{sec:LO}), whereas phonons with much lower frequencies are unable to respond effectively to such changes. Thus, even though the polaron transformation does not explicitly make presumptions about the frequencies of the bath and the TLS, the variational procedure automatically incorporates a time scale, associated with renormalized TLS tunneling rates, that captures the essential features of the response of both slow and fast bath variables. 

\subsection{Energy relaxation} 

As mentioned in Appendix \ref{sec:polaron}, the polaronic TLS state is a strongly correlated state of the unrenormalized TLS and  phonon bath, in which the phonon displacements are correlated with the TLS's internal state. The optimal values of $\alk$ and $\bk$ describe the mean shifts of phonons in correlation with the TLS, and the residual interaction with the bath, represents fluctuations beyond this mean field shift that has vanishing vaccum expectation values. The polaronic TLS, while encapsulating all the strong correlations of the unrenormalized TLS and phonon bath, is itself only weakly correlated with the bath. Thus, one can now apply the Born-Markov approximation (see Appendix \ref{sec:a} for more details) to derive a master equation describing the decoherence properties of the polaronic TLS and attain results for the relaxation and dephasing times of the TLS. 

 In particular, for relaxation, one finds that even though ostensibly Eq. (\ref{eq:final}) suggests a different rate of decay due to change of $g_k$ to $g_k - \alpha_k$, following the Born-Markov procedure results in a (single phonon) decay rate that is $exactly$ the same as that obtained from the Fermi Golden Rule approach applied to the Hamiltonian with unrenormalized phonon coupling (but with renormalized values of $E_z$ and $\Delta$) prior to the polaron transformation. Further, we find that all multi-phonon processes in which the TLS decays by releasing more than one phonon, are completely insignificant in the temperature range of experimental interest, and do not affect the decay rate. It must also be mentioned that TLS-TLS interactions, which will be studied in more detail in Section \ref{sec:intdec}, statistically (as very few are expected to be resonant with a given TLS) contribute insignificantly in comparison to direct phonon mediated decay, but when a single TLS's energy is tuned, resonances may occur and TLS induced decay may become significant.  The effective relaxation rate $T^{-1}_1$ of the excited state to the ground state can then be found (as known from Fermi's Golden rule \cite{Phillips})
\be
T^{-1}_1 = \sum_{\alpha} (2 \pi)^{3} \frac{\gamma_z^{2}}{ v_{\alpha}^{5}} \; \frac{(2E)^{3}}{\rho h^4} \, \sin^2 \theta \; \coth{ \frac{E}{k_B T}}
\label{eq:relax},
\ee 
where $\sin \theta = \Delta/ E$ and $\alpha$ represents the various phonons modes that couple to the TLS. For a TLS of energy 8 GHz, $\gamma_z \sim 1$ eV, this implies a decay time $T_1$ of about $135$ ns $/ \sin^2 \theta$. For the data on the TLSs presented in \cite{Ustinov}, with $T_1$ times of about 400 ns, implies that $\sin \theta = \Delta/E \sim 0.5$. It should be noted that it is, in fact, necessary for this factor of $\sin \theta$ to be significant because the same factor precisely enters into the Qubit-TLS coupling strength upon diagonalization in Eq. (\ref{eq:q}) and is thus crucial for the TLS to produce a visible gap in the qubit spectrum. 

\begin{figure}
\begin{center}
\includegraphics[width = 3.5 in]{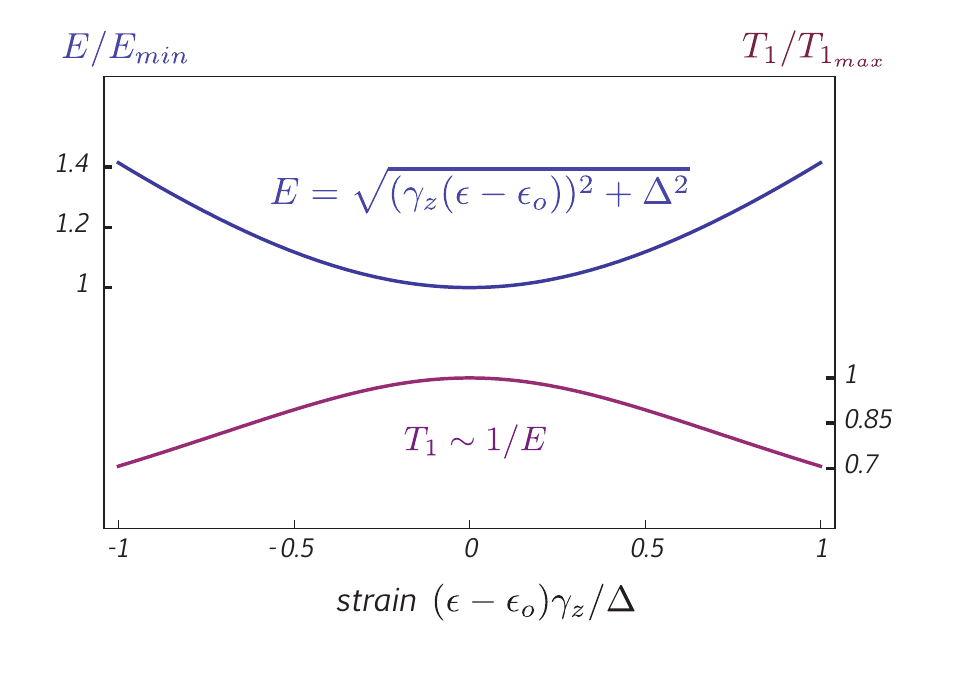}
\end{center}
\caption{Schematic showing the change in energy $E$ and relaxation time $T_1$ of the TLS as uniform strain $\epsilon$ applied on the system is varied. The Energy $E$ is minimal ($E = E_{min}$) for strain $\epsilon = \epsilon_o$. At this value of the strain, $T_1$ concurrently reaches its maximum value $T_{1_{max}}$.}
\label{fig:strain}
\end{figure}

We would now like to point out an intriguing aspect of this phonon induced decay rate. Since $\Delta$ remains roughly constant as the strain is varied (see section \ref{sec:AC}), one can glean from Eq. (\ref{eq:relax}) that the decay rate is actually directly proportional to $E$ (recall $\sin^2{\theta} = (\Delta/E)^2$). This implies that as the energy of the system is tuned using the strain to its minimum value, the lifetime of the TLS actually increases to its maximum value (see Fig. \ref{fig:strain}). This result is unusual because in principle our expectation would be for a symmetric double well to be more susceptible to decay due to noise that tampers with this symmetry than an already asymmetric double well (cf. \cite{relaxation}). In mathematical terms, the matrix element that accompanies the decay rate is $\lrs{\Delta/E}$ which clearly implies that decay rate must increase as the energy $E$ is lowered. What reverses this trend is the fact that the phonon density of states also goes as $E^2$ besides the factor of $E$ that comes from matrix element of the strain field. Thus, an increase of $T_1$ as the TLS is tuned to its energy minimum is really due to the super-ohmic nature of the strain field responsible for TLS decay. This can be a simple but effective signature for experiments to check for phonon induced decay of TLSs (see also the discussion in Section \ref{sec:intdec} of the effect of TLS-TLS interaction on this result.)

While the above analysis holds true for a single TLS whose energy is being modulated by applying strain, a similar comment can be made about the distribution of the TLSs $T_1$ times vs.\ energy $E$. When calculating such a distribution, one should be careful to interpret Eq. (\ref{eq:relax}) as really- 

\be
T^{-1}_1 = \sum_{\alpha} (2 \pi)^{3} \frac{\gamma_z^{2}}{ v_{\alpha}^{5}} \; \frac{8 E \Delta^2}{\rho h^4} \; \coth{ \frac{E}{k_B T}} . 
\label{eq:relaxm}
\ee 

In principle, we should integrate over the distribution of $\Delta$ to find the true distribution of TLSs $T_1$ vs.\ $E$. As we mentioned earlier, we expect $\Delta$ to be usually smaller or at best comparable to the asymmetry energy $E_z$. Under this assumption, $\Delta$ does not affect $E$ significantly, and we thus expect the scaling of the typical relaxation time $T_1$ of TLSs at a given energy $E$ to go as $1/E$, although we expect strong variations due to random values of $\Delta$. 

\begin{figure*}[t]
  \subfloat[]{\label{fig:2a}\includegraphics[width=4.3in]{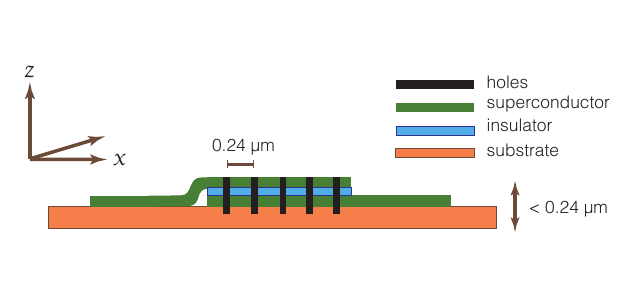}}                
  \subfloat[]{\label{fig:2b}\includegraphics[width=3in]{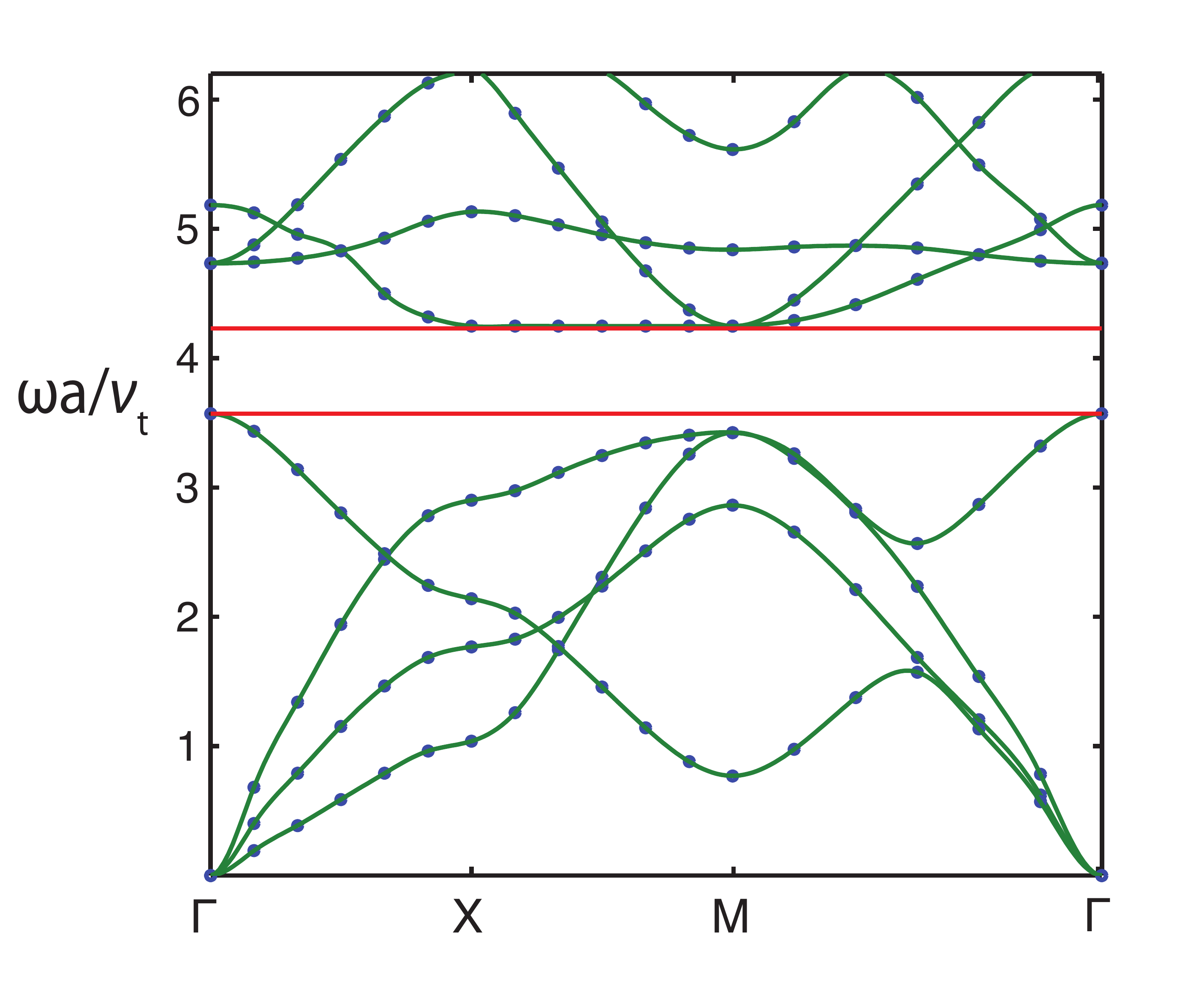}}
  \caption{(a) The proposed design of the qubit. Note that the holes are drilled in a square lattice pattern at regular intervals of 0.24 $\mu$m and the qubit must be fabricated on a substrate no greater than the same length, (b) The band structure for $k_z = 0$. $\Gamma$, X, and M represent the points ($k_x a$,$k_y a$) = (0,0), (0,$\pi$),($\pi$,$\pi$)respectively. The band gap extends from $ \omega a / v_t =  3.57$ to $ \omega a / v_t =  4.23$. }
  \label{fig:2}
\end{figure*}

\subsection{Decoherence and dephasing}

Pure dephasing due to phonons can be found from the Born-Markov approach and the result is mentioned in Eq. (\ref{eq:t2}). At its largest, the dephasing rate is found to be of the order of $1$ KHz which is still completely negligible in comparison to the decay rate. The pure dephasing rate can however also be estimated in a more direct and appealing fashion - by estimating the effect of the extra phase difference $\phi (t)$ between the eigenstates of the TLS generated from the interaction with the phonons -- 
\be
\phi (t) = 2 \int_{0}^{t} \sum_k \; \gamma_{l,k} \epsilon_k(t). 
\ee
Here $\gamma_{l,k} \epsilon_k = (E_z/E) (g_k - \alk) ( \ak + \ad) $, and $\gamma_{l,k}$ has a $k$ dependence due to the presence of $\alpha_k$ (the subscript $l$ is to indicate the extra factor of $(E_z/E)$ in comparison to $\gamma_z$). Using the assumption that the strain field has a gaussian distribution implies that $< e^{i \phi(t)}> = e^{-<\phi(t)^{2}>/2}$. Further, assuming that the bath of phonons is uncorrelated with the TLS and in thermal equillibrium, we arrive at the result 
\be
\avg{\phi(t)^{2}} = \int_0^{\omega_D} d \omega \; \frac{\omega}{ \rho \hbar v_t^2} \frac{4 \gamma_{l,k}^{2}}{ \pi^2 v_t^3} \coth{\left( \frac{\hbar \omega}{2 k_B T} \right)} ( 1- \cos{\omega t} ) .
\label{eq:dephase}
\ee

We find that the time dependent part has an extremely small amplitude of the order of $10^{-4}$, and performs underdamped oscillations of the form $ \sin(\omega t) / t$. Further, the time independent part is also small (of the same order as the time dependent part). Therefore, the phonons indeed do not cause any significant pure dephasing, which allows for the common observation of TLSs with a decoherence time $T_2 \sim 2 T_1$ in experiments \cite{Ustinov}. It is also worthwhile to note that if we ignore $\alk$ in Eq. (\ref{eq:dephase}), the time independent part of this integral is the essentially the same as that in the exponent of the factor that renormalized $\Delta_0$ in Eq. (\ref{eq:det}) (setting $\alk \sim g_k$ for high frequency phonons). Thus, the polaron transformation acts by effectively shifting the effect of constant depahsing by high frequency phonons to that of renormalizing the tunneling strength. We also already know the estimate of this integral to be of the order of $10$, which implies that prior to this process of systematically addressing the role of high frequency phonons in TLS renormalization, we would have falsely concluded that the TLS decoheres extremely (anomalously such that near instantly, the off-diagonal matrix element gets reduced by a factor $e^{-10}$) rapidly.

Finally, we would like to point out that these estimates of decoherence are due to interaction with phonons only. We do not always expect TLSs to have $T_2 \sim 2 T_1$ as there will be, in general, other sources of decoherence (such as the TLS-TLS interactions analyzed in Section \ref{sec:intdec}) coupled to the TLS, making $T_2$ shorter. 


\subsection{Phononic band gap engineering}
\label{sec:crystal}

Since TLS decoherence is strongly influenced (with $T_2$ being of the order of $T_1$) by relaxation processes, one may expect to get a significant increase in coherence times if we are able to increase the lifetime of the TLS excited state. It has been demonstrated that due to their favorable decoherence and relaxation properties, the TLSs can be used themselves as {\em accidental} qubits \cite{qmem}. Therefore, getting longer coherence times could enable new types of quantum-computing related experiments. Since we find that phonon-induced decay largely explains the relaxation rate, the most obvious way of achieving longer lifetimes for TLSs is to open up a gap in the phonon spectrum by drilling holes at regular intervals into the Al superconducting device. Such phononic crystal structures have in fact, been recently experimentally realized \cite{hypersonic}. We find that a reasonable gap of about 2 GHz can be opened up near the 8 GHz frequency mark\footnote{We use FEM to perform band structure calculations. These are in fact in excellent agreement with results found using the FDTD method \cite{Alstudy}. See also \cite{oblique}, and \cite{FEMboundary} for details on FEM based band structure calculations} (Fig. 2b illustrates the band gap in terms of dimensionless parameters), by drilling holes at intervals of $a = 0.24 \mu m$ in a square lattice shape, and with a filling fraction $f = 0.60$. This gap however, exists primarily for phonons being radiated in the x-y plane (see Fig. 2a). Thus, TLSs can still radiate in many other directions even with the phononic crystal structure in place. To make the TLS truly immune to decay via phonons, we propose that the fabrication of the entire device be performed on a thin film no greater than $0.24 \mu m$ in thickness. This procedure renders the phonon of the wavelengths of interest (those that cause TLS decay) to become effectively 2D; they now reside in a complete band gap, and are in principle, not susceptible to decay via phonons. 

As we mentioned earlier, pure dephasing should go to negligibly small values (as observed in experiments \cite{unpub}) at the symmetry point ($E_z = 0$) for TLSs, as their energy is tuned by applying strain. At this point, decoherence from all noise sources, is almost entirely  relaxation driven (of which we believe phonons to be the major source of energy relaxation), and such an experimental scheme may prove to be useful in dramatically enhancing TLS coherence times.  

\section{TLS-TLS interactions and decoherence}
\label{sec:intdec}

Since we find TLS pure dephasing due to phonons to be negligible, the role of TLS-TLS interactions in dephasing and perhaps even relaxation must be addressed. TLS-TLS interactions have been studied in previous works \cite{BH,Paris,Ioffee}. In particular, Black and Halperin \cite{BH} found that by treating the problem classically, neglecting the dynamics of the TLSs, one can arrive at an expected phonon strain\footnote{TLS-TLS interactions mediated by the electric dipole field are also of the same form, and coincidentally, of the same order in magnitude} mediated dipole-dipole like interaction between the TLSs of the form (in the TLS eigenbasis) --
\be
V_I = \frac{C_{ij} (\hat{r}_{ij})}{4} \frac{1}{r_{ij}^3} \left( \frac{E^z_i}{E_i} \sigma^z_i + \frac{\Delta_i}{E_i} \sigma^x_i  \right) \left(  \frac{E^z_j}{E_j}  \sigma^z_j + \frac{\Delta_j}{E_j} \sigma^x_j \right) .
\label{eq:BH}
\ee
where $i$ and $j$ are indices of interacting TLSs, and $C_{ij}(\hat{r}_{ij})$ is a material-specific constant that depends only on the direction of the vector connecting TLSs $i$ and $j$. To provide for an estimate, Black {\em et al.} \cite{BH} perform $rms$ angular averaging over the direction $\hat{r}_{ij}$, and find for typical values of TLS-phonon interaction, phonon velocities etc., $C_{rms} \sim 1.6 \times 10^{-42} J cm^3$. We will use this result of Black {\em et al.} \cite{BH} as the basis for discussing decoherence of TLSs, due to TLS-TLS interactions. For completeness of this paper, and to establish a connection between the polaron approach and previous works, in Appendix (\ref{sec:TLSinteraction}), we show how this interaction follows from the polaron transformation procedure in the appropriate limit. 

For the contribution of other TLSs to the decay rate of a given TLS, we need to estimate the average interaction $V^{res}_{I}$ of our chosen TLS with near resonant TLSs. The energy window in which we consider TLSs to be resonant is given by max ($1/T_1$,$V^{res}_{I}$), where $T_1$ is the relaxation time of our chosen TLS due to phonons. We take for example, a TLS of energy $2E\sim 10$ GHz, and $T_1 \sim 1 \mu s$ as found earlier. Using the density of states $\nu \sim 10^{45} J^{-1} m^{-3}$, and assuming $1/T_1 \gg V^{res}_I$, the spatial density of TLSs in resonance is given by $\rho_{res} = \nu (1/T_1)$, and consequently the average interaction strength $V^{res}_I \sim C_{rms} \rho_{res} = C_{rms} \nu (1/T_1) \sim 0.5$ KHz. Thus our assumption $1/T_1 \gg V^{res}_I$ is justified. The decay rate then, is given by $(V^{res}_I)^2 S_X (\om = 2E_i)$ where $S_X$ is the correlator of the bath of TLSs that couples to the $\sigma^x_i$ term of our TLS $i$ in Eq. (\ref{eq:BH}). For large frequencies, $S_X (\om)$ is primarily given by the $\avg{[\sigma^x_j(\om), \sigma^x_j(-\om)]_+}$ correlators \cite{Ivarm} of the bath TLSs, that are Lorentzians centered at $\om = 2 E_j$ with width $1/T_{2,j}$, where $T_{2,j}$ is the decoherence time of TLS $j$. For the case of resonant TLSs, $E_j \sim E_i$, the decay rate is given by $ (V^{res}_I)^2 T_{2,j} \sim 10^{-3}$ KHz which is significantly smaller  than the contribution to decay by phonons. Here we used the experimental observation that for most TLSs in the energy range of $10$ GHz, as our example TLS, $T_{2,j}$ is of the order $T_{1,j} \sim 1/$MHz. Thus, statistically, we expect that TLS decay due to other TLSs is negligible. Nevertheless, we would like to point out that when a selected TLS's energy is tuned by applying strain, it is bound to come into resonance with some TLSs. Coupling between such TLSs depends on the distance between them, so it should have strong variations from one resonance to another. At such TLS resonances we expect to find considerable variation in the $T_1$ in the range of strain, where the difference of TLS energies is comparable to their coupling. These rapid resonance variations  of the relaxation rate can make it difficult to verify the smooth trend $T_1\propto 1/E$ profile that we obtained earlier.

We now look at the contribution of other TLSs in the dephasing of our TLS. We expect only TLSs that are thermally active (of half energy $E_j$ below the temperature $T$) to contribute to dephasing. Assuming that the temperature of the system is lower than the energy of the TLS we are examining (under typical experimental conditions $T \sim 100$ mK is a few times smaller than TLS energy $2 E \sim 10$ GHz), these thermally activated TLSs are expected to be slower than our TLS (recall from Eq. (\ref{eq:relaxm}), decay times typically decrease with decreasing TLS energies). The dephasing due to such slow TLSs can be obtained only after averaging over many experimental runs, each of which would correspond to a different but near static configuration of the bath. The flipping of such slow TLSs thus, merely shifts the energy of our TLS, at the scale of the typical interaction energy $V^{non-res}_I$. As we discuss below, dephasing due to such TLSs is qualitatively different from a bath that operates at time scales faster than our TLS.

For the general case of a gaussian bath, one finds the pure dephasing induced decay of the off-diagonal density matrix element of the TLS is given by \cite{slownoise}
\be
 \avg{\sigma^-_i (t)} = \exp \left( -\frac{1}{2} \int \frac{d \om}{2 \pi} S_Z (\om) \frac{\sin^2 (\om t/2)}{ (\om/2)^2} \right) , 
 \label{eq:deph}
\ee
where $S_Z(\om)$ is the symmetrized bath correlator in the Fourier domain, $S_Z(\om) = 2 \int^{\infty}_{-\infty} dt e^{i\om t}  \avg{[V_z (t) , V_z (0)]_+} $. Here, $V_z$ is sum total of the interaction terms from all TLSs $j \ne i$ that couple to the $\sigma^z_i$ operator of TLS $i$ in Eq. (\ref{eq:BH}). At low frequencies, the major contribution to $S_Z (\om)$ comes from the correlators  $\sum_j \avg{[\sigma^z_j(\om), \sigma^z_j(-\om)]_+}$ that are given by \cite{Ivarm} 
\be
\avg{[\sigma_z^j(\om), \sigma_z^j(-\om)]_+} = \frac{2 T_{1,j}}{1 + \om^2 T^2_{1,j}} \frac{1}{(\cosh{E_j/k_bT})^2}.
\ee 
Thus, only TLSs with half energy $E_j < k_b T$ contribute to dephasing, which is why we consider only thermally activated TLSs. Given the $1/r^3$ nature of the interaction between TLSs, nearest neighbor TLSs are expected to most strongly affect TLS decoherence properties, and make a dominant contribution to the net bath correlator $S_Z (\om)$. Motivated by these observations, to bring out the difference between pure dephasing due to a slow bath and a fast one, we consider a bath with correlator $S_Z (\om) = A^2 \Gamma/(\Gamma^2+\om^2)$, which is a telegraph noise with a switching rate $\Gamma$ and interaction strength $A$, and we evaluate the exponent in Eq. (\ref{eq:deph}) in the two regimes, $1/\Gamma \gg t$, appropriate for a slow bath and $1/\Gamma \ll t$ for a fast bath, where the time $t$ that we're interested in, is set by the time scale at which the TLS under examination decoheres (say around $1 \mu s$). Now, the scale at which the integral decays in the exponent of Eq. (\ref{eq:deph}) is set by either $S_Z (\om)$ which decays at the scale $\Gamma$ or $(\sin^2 (\om t/2))/\om^2$ that decays at the scale $1/t$. For a short-correlated bath, we find $\avg{\sigma_- (t)} \sim \exp \left( - (A^2/\Gamma) t \right)$. For a slow bath, with $\Gamma \ll 1/t$, we find instead $\avg{\sigma_- (t)} \sim \exp \left( -A^2 t^2 \right)$ wherein the only time scale is given by $1/A$. In the present case, we are in the regime where the TLS bath's switching rate, set by $\Gamma \sim 1/T_{1,j} \ll 1/t \sim 1 \mu s$. That is, the TLS bath is slower than the TLS under examination.

Thus, in the case of dephasing of a given TLS by other TLSs, the only scale that decides the additional contribution to dephasing is given $A$, or in this case, $V^{non-res}_I$ and the decay is gaussian, that is, $\exp \left(-(V^{non-res}_I)^2 t^2 \right)$. To estimate $V^{non-res}_{I}$, we first note that the density of thermally activated TLSs is given by $\rho_{non-res} \sim \nu T$, where $T$ is the temperature and, consequently $V^{non-res}_I \sim C_{rms} \nu T \sim 0.5$ MHz, at $T = 100 mK$. Since this is of the order of linewidth of our TLS, we expect strong variations in the values of $1/T_2$ whereby the nearest neighbor thermally active TLS strongly influences the decoherence property of a given TLS. When this nearest neighbor distance is shorter than the typical distance $r_{typ} \sim (1/\nu T)^{1/3} \sim 100$ nm for $T = 100 mK$, we expect interactions with TLSs will dominate the dephasing of our TLS. However, if this distance is larger than $r_{typ}$, we expect TLS decoherence to be set by phonons, with $T_2 \sim 2 T_1$. Nevertheless, from the above analysis, it seems reasonable to expect, that for most TLSs of energies of the order of $10$ GHz, $T_2$ will be at least of the same order as $T_1$, which is what one finds in experiments.  

To complete this discussion, we point out that the different nature of decoherence due to phonons (via decay) and TLS-TLS interactions (via pure dephasing) allows us to simply add up their corresponding contributions to the decay of the off-diagonal term of the TLS density matrix. In a single experimental run, the TLS bath shifts the frequency of our TLS slightly, simply modifying oscillation frequencies in a Ramsey experiment. The gaussian form of decay due to TLS-TLS interactions appears upon averaging over many experimental runs. Hence, we suggest that to analyze such experimental data, one should carefully fit the decay with a more general 2 parameter fit of $\exp{\left(-A^2 t^2 - B t \right)}$, where $B$ should be approximately $1/(2 T_1)$. It is also important to note that the interaction $A = V^{non-res}_I \propto (E^z/E)_i$ for the TLS $i$ under observation. This means that at the point of TLS energy minima ($E^z_i = 0$, when tuned by applying strain), one should achieve minimal dephasing. At this point, decoherence is truly driven by phononic relaxation, that can be minimized via the fabrication scheme described in Section \ref{sec:crystal}. 

Again, we remind our reader that this tuning of pure dephasing to zero at the symmetry point of the TLS holds very generally in our theory. Since all the coupling of the environment to the TLS via its $\sigma_x$ operator is strongly suppressed (by the same large factor that the tunneling is suppressed by), the TLS's coupling to the environment is expected to be primarily through its dipole operator, represented by $\sigma_z$ in the charge basis. Consequently, the interaction of the TLS with the environment in its eigenbasis must of the form -- $(E_z/E) \sigma_z + (\Delta/E) \sigma_x$, which means that as we approach the symmetry point of the TLS, where $E_z = 0$, the pure dephasing rate becomes entirely negligible, yielding a decoherence time $T_2 \rightarrow 2 T_1$ at this point. As far as we know, experiments \cite{unpub} are in complete agreement with this statement. 

\section{Conclusions and Outlook}

We conclude with the understanding that tunneling systems observed in bulk amorphous solids, as propounded by Anderson {\em et al.} \cite{Anderson} and Phillips \cite{Philold} are actually electrons dressed by optical frequency phonons tunneling between atoms rather than atoms or group of atoms tunneling in double well structures. This reasoning is motivated by the fact that coupling strengths suggested by experiments both on bulk solids and in JJ Qubits  are too large to ignore their impact on the renormalization of the TLS parameters. Further, once we assume that it is indeed an electron that is the tunneling object, explanation of the coupling to acoustic phonons becomes transparent, the density of such TLSs can be estimated to the correct order of magnitude as seen in experiments, and the coupling of such TLSs to the JJ Qubit is also easily explained. 

Due to renormalization of TLSs by optical frequency phonons, the decoherence and relaxation properties are also significantly altered, and in particular pure dephasing due to phonons becomes insignificant. TLS dephasing is then given primarily by interaction with other TLSs, while the low temperature relaxation rate is given primarily by phonon induced decay. In particular, we find that pure dephasing due to the bath of TLSs leads to a gaussian in time decay of the off-diagonal density matrix element, with an effective time scale given by the average interaction of the TLS with other thermally activated TLSs in the bath. For $T= 100$ mK, we find this time scale for pure dephasing to be of the order of the decay rate (due to phonons) of TLSs of energy in the GHz range, explaining quantitatively, why for most TLSs at such energies, experiments find $T_2$ to be of the order of $T_1$ if not twice, with $T_1 \sim 1 \mu$s \cite{Ustinov}. 

We also provide an explanation for the existence of a minima in the energy of the TLS as a function of strain, and predict that phonon induced decay must produce the intriguing signature of a maximum of relaxation time $T_1$ at this symmetry point contrary to usual expectations. The renormalization procedure we present also has the important consequence that any coupling to the environment must be through the dipole ($\sigma_z$) operator in the TLS charge basis. This directly leads to the practically important result that the dephasing rate of the TLS must become insignificantly small as it is tuned to its symmetry point by the application of stain and explains why experiments consistently observe that $T_2 \sim 2 T_1$ at this symmetry point \cite{unpub}. An important outstanding issue is that of explaining the temperature dependence of the decay rate of the TLS; we believe this to be a result of coupling not only to phonons, but also to the other parts of Qubits environment, including non-equilibrium quasiparticles \cite{Glazman} in the superconducting Al layer. 


Finally, we presented a schematic of an experimental setup that should result in suppressing phonon-induced decay of TLSs, enhancing their lifetime, and consequently decoherence times, significantly. This experimental scheme becomes more viable as TLSs are tuned to their energy minimum as strain is varied, since we expect dephasing by other TLSs at this symmetry point to be minimal. While we don't expect this to improve the coherence properties of superconducting devices themselves, (for whom these TLSs are a major source of decoherence \cite{Ivarm,Spin,noise,Martinis04}), this scheme could have an important impact on not only improving the viability of TLSs for use as quantum memories \cite{qmem}, but should also allow us to understand better the other sources of noise that decohere the TLSs, once the influence of phonons is significantly reduced. A better understanding of these aspects of the TLS should eventually help improve the practicability of both superconducting devices and optomechanical resonators \cite{Kippenberg} where these TLSs significantly limit the coherence times and Q-factors respectively, as well as engineer exciting new nano-mechanical devices with strong non-linearities of phonon modes.  


\section{Acknowledgements}

We thank B.I. Halperin, A. Ustinov, J. Lisenfeld, A. Shnirman, L.B. Ioffe, A. Amir, V. Manucharian and G. Refael for useful discussions. We would like to especially thank A. Ustinov for sharing his unpublished data. KA likes to thank M. Babadi for insightful comments. Work at LANL was carried out under the auspices of the National Nuclear Security Administration of the U.S. Department of Energy at Los Alamos National Laboratory under Contract No. DE-AC52-06NA25396 and supported by the LANL/LDRD Program. The authors acknowledge support from a grant from the Army Research Office with funding from the Harvard-MIT CUA, NSF Grant No. DMR-07-05472, ARO MURI Atomtronics and ARO MURI Quism programs. 

\appendix

\section{TLS-acoustic phonon coupling induced by phonon nonlinearity}
\label{sec:grun}
The origin of Gruneizen parameter lies in non-linear phonon coupling,  $H_{NL} = \gamma x^2 \epsilon(r) $. The form of this term is based of the following intuition -- the first generic non-harmonic term has to be cubic in the phonon fields; it must involve two optical phonon field terms whose frequencies nearly cancel each other so as to give a resultant acoustic frequency, and that the coupling to acoustic phonons must be via a derivative of the acoustic phonon field. In effect, $H_{NL}$ serves to modify the stiffness constant $k$ of the optical phonon mode, which allows us (using the definition of the Gruneizen parameter $\eta = d log \om / d \epsilon$, $k = m \om^2$) to arrive at the following relation -  $\gamma = \eta k$. Thus, we can write the following Hamiltonian   
\begin{align}
H_o + H_{op} + H_{NL} &= E_z^0 \sz + \Delta_0 \sigma_x +  \nonumber \\
& \frac{p^2}{2m} +  \frac{1}{2} k \left(1 + \eta \epsilon(r) \right) x^2 + \lambda x (\sz + c).
\end{align}
 
Eliminating the optical phonon mode now follows the same procedure as before -- the only difference being a modified value for $k$, and we finally obtain
 \begin{align}
 H_{TLS} + H_{ac} &= E_z^0 \sz + \Delta \sigma_x  - c E_p / \left(1 +  \eta \epsilon(r) \right) \sz.
 \label{eq:nonlin}
 \end{align}
 
Expanding the above in $\eta \epsilon(r)$ gives us the desired acoustic phonon-TLS coupling, and hence reproduces the Hamiltonian in Eq. (\ref{eq:1}). Since $\eta \epsilon (r) \ll 1$ (at $T = 100$mK, $\avg{\sqrt{\epsilon^2(r)}} < 10^{-8}$), all terms beyond the first order in $\epsilon (r)$ can be safely neglected. 

\section{Polaron Transform}
\label{sec:polaron}

We begin with the general Hamiltonian coupling a TLS and a bath of phonons as in Eq. (\ref{eq:H}) and apply to it the polaron transformation $U$ mentioned in Eq. (\ref{eq:trans}). Then $\tilde{H} = U^{\dagger} H U $ is given by 
\begin{align}
\tilde{H} &= E'_c + E'_z \tilde{\sigma}_z + \Delta \tilde{\sigma}_x + \sum_k \; ( \ak + \ad ) \; \tilde{\sigma}_z \; ( g_k - \alk )  \nonumber \\
& \sum_k \ok a^\dagger_k \ak  \nonumber + \Delta_0 \tilde{\sigma}_+ \left( e^{-2 \alpha} - \avg{e^{- 2 \alpha} } \right) \\
&  + \Delta_0 \tilde{\sigma}_- \left( e^{2 \alpha} - \avg{e^{ 2 \alpha} } \right) + \sum_k ( \ak + \ad ) ( c_k g_k - \bk).
\label{eq:Ham1}
\end{align}
Here $\tilde{\sigma}$ are the pauli matrices in the $U$-transformed basis. The values of $E'_c$ and $E'_z$ are
\begin{align}
E'_c &=  \sum_k \left( - 2 \frac{\bk^* c_k g_k}{ \ok } -  2  \frac{\alk^* g_k}{\ok} + \frac{\abs{\alk^2}}{ \ok} + \frac{\abs{\bk^2}}{\ok} \right) ,\nonumber \\
E'_z &= E^z_o - \sum_k \left( 2 \frac{\bk^* g_k}{ \ok } +  2  \frac{\alk^* c_k g_k}{\ok} - 2 \frac{\alk^* \bk}{\ok} \right) , \nonumber \\
\Delta &= \Delta_0 \avg{e^{2 \alpha}} , \nonumber \\
\alpha &=  \sum_k \alpha_k/\om_k (a_k - \ad) .
\label{eq:ec}
\end{align}


Note that the reality of these terms is guaranteed by the conditions on the hermicity of the Hamiltonian and the unitarity of the polaron transformation (which requires, $\alk = \alpha^*_{-k}$, and $\bk = \beta^*_{-k}$). Now, from the form of $U$ in Eq. (\ref{eq:trans}), one can see that the polaron transformation is simply a unitary transformation that allows us one to go from a basis of independent TLS ($\sigma$) and phonon variables, to the new TLS variable ($\tilde{\sigma}$) that corresponds to a correlated motion of phonons and the unrenormalized TLS. The philosophy then is to somehow optimize these correlations such that the free energy of the polaronic TLS system, which now is only weakly correlated to the residual bath, is minimized. To do so, one introduces the set of parameters $\alk$ and $\bk$, that describe the amplitude of phonon displacements correlated with the TLS's internal state and provide a variational space over which to minimize the free energy of the polaronic TLS. The optimal values of $\alk$ and $\bk$ thus, represent the mean phonon displacements in correlation with TLS state, and the residual coupling with the bath represents fluctuations about this mean displacement, that has vanishing vacuum expectation values. 

Given the polaron-transformed Hamiltonian $\tilde{H}$ in Eq. (\ref{eq:Ham1}), we need to estimate the minimum of the free energy. But first, we must complete the squaring of the phonon field to eliminate the last term in Eq. (\ref{eq:Ham1}). This is brought about by the transformation - $\tilde{a}^{\dagger}_k = \ad  - (\bk - c_k g_k)/ \ok$. In fact, this transformation eliminates $\beta_k$ (replacing it with $c_k g_k$) from the transformed Hamiltonian completely leaving us with a variational state dependent only on $\alpha_k$. It is important to note that this last transformation leaves the coefficients of $\tilde{\sigma}_+$ and $\tilde{\sigma}_-$ unchanged in Eq. (\ref{eq:Ham1}) and these can now be treated as perturbations to the phonon and TLS parts of the Hamiltonian, as their expectation values in the thermal ensemble of phonons is zero. In the process of separating these these final terms, we arrived at the mean field estimate for the renormalized tunneling, $\Delta = \Delta_0 \avg{e^{2 \alpha}}$  .

To perform the approximate minimization of the free energy, we minimize the Bogoluibov-Peierls upper bound on the free energy (which here is simply the free energy of the TLS and phonons decoupled from each other) $F = E_c - (1/\beta) lnZ + F_{ph}$. $Z$  is partition function of the TLS, $e^{-\beta E'} + e^{\beta E'}$ with $E' = \sqrt{E'^2_z + \Delta'^2_o} $, and $F_{ph}$ is the free energy of the phonons which is independent of $\alpha_k$. The result of this is the following equation on $\alk$, 
\begin{equation}
 (\alk - g_k)  + \left( \frac{\Delta}{E} \right)^2  \frac{2 E}{\ok} \alk \tanh{\beta E} \coth{\beta \om_k /2} = 0 .
 \label{eq:min}
\end{equation}

Solving this leads to the main results in Eq. (\ref{eq:det}). Also, note that since $g_k$ is real for this case, so is $\alpha_k$ and we consequently drop the usage of extra symbols such as $*$ and $\abs{}$. 

\section{Calculation of decay rate}
\label{sec:a}

The first step in this procedure is to diagonalize the TLS part of the Hamiltonian in Eq. (\ref{eq:final}). The Hamiltonian can then be represented by
\be
H_o = E \tilde{\sigma_z} + \vd \tilde{\sigma}_{+} + \vi \tilde{\sigma}_{-} + \vo \tilde{\sigma}_z , 
\ee
where
\begin{align}
\vo &= \zc \sum_k (g_k - \alk) (\ak + \ad ) + \frac{\deo}{2} \xc X + \frac{\deo}{2} \xc \xd , \nonumber \\
\vd &= \frac{1}{2} \left( \zc + 1 \right) \deo X + \frac{1}{2} \left( \zc - 1 \right) \deo \xd \nonumber \\
 & - \xc \sum_k (g_k - \alk ) ( \ak + \ad), \nonumber \\
 X &= ( e^{-2 \alpha} - \left<e^{- 2 \alpha} \right> ) . \nonumber \\
\end{align}
The time evolution of the density matrix in the Born-Markov approximation \cite{zzz} is described by the equation --
\be
\d{\rho}{t} =  - \int_0^{\infty} ds Tr_B \left[ V_I (t) , [ V_I (t-s) , \rho_S (t) \otimes \rho_B ] \right] .
\label{eq:a3}
\ee

Expanding the RHS of Eq (\ref{eq:a3}) out, and keeping only the slowly fluctuating terms (the rotating wave approximation), we arrive at the estimates for $T_1$ and $T_2$ -- 

\begin{align}
\frac{1}{T_1} &= \int_{0}^{\infty} ds e^{i2Es} \lrb{\vd(s) \vi(0)} + h.c. \nonumber \\
&+ \int_{0}^{\infty} ds e^{-i2Es} \lrb{\vi(s) \vd(0)} + h.c. , \nonumber \\
\frac{1}{T_2} &= \frac{1}{2 T_1} + 4 Re\left[ \int_{0}^{\infty} \lrb{\vo(s) \vo(0)} \right] .
\end{align}

More explicitly,  
\begin{align}
\frac{1}{T_1} &= 2 \pi \lrs{\xc} \sum_k \lrs{g_k - \alk} \lr{ 2 n_k + 1} \delta ( \ok - 2E) \nonumber \\
&+ 8 \pi \lr{\xc} \dea \sum_k \aok \lr{g_k - \alk} \lr{ 2 n_k + 1} \delta ( \ok - 2E) \nonumber \\
&+ \dea^2 \lr{ \lrs{\zc} + 1} Re \left[ \int_{0}^{\infty} e^{i2Es} ds \lr{ e^{G(s)}   -1}  \right] \nonumber \\
&+ \dea^2 \lr{ \lrs{\zc} - 1} Re \left[ \int_{0}^{\infty} e^{i2Es} ds \lr{ e^{ -G(s)}   -1}  \right] \nonumber \\
&+ \dea^2 \lr{ \lrs{\zc} + 1} Re \left[ \int_{0}^{\infty} e^{-i2Es} ds \lr{ e^{ G(s)}   -1}  \right] \nonumber \\
&+ \dea^2 \lr{ \lrs{\zc} - 1} Re \left[ \int_{0}^{\infty} e^{-i2Es} ds \lr{ e^{ -G(s)}   -1}  \right]  , 
\label{eq:t1}
\end{align}

\begin{align}
\frac{1}{T_2} &= \frac{1}{2 T_1} + 2 \Delta^2 \lrs{\frac{\Delta}{E}} Re \left[ \int_{0}^{\infty} ds \left( e^{-G(s)} + e^{G(s)} - 2 \right) \right] ,
\label{eq:t2}
\end{align}
where,
\be
G(s) = 4 \sum_k \lrs{\aok} \nbt .
\ee

Adding up the processes involving only 1 phonon operator in Eq. (\ref{eq:t1}) gives the result in Eq. (\ref{eq:relax}). As for the pure dephasing rate, $\Gamma_{\phi} = 1/T_2 - 1/2T_1$ being negligible, a quick way to see this is to realize that if we expand the exponentials in Eq. (\ref{eq:t2})), the first non vanishing term comes with two Bose distribution functions, proportional to $n (\om) ( n (\om) + 1)$ (with the same frequency)  which decays  rapidly for frequencies at which $\alk$ is appreciable. 

\section{TLS-TLS interactions using Polaron Transformation}
\label{sec:TLSinteraction}

To get a full description of the TLS-TLS interaction, one needs to include the full tensorial structure for the strain field that couples to the TLS. In particular, one has $\epsilon_{\alpha \beta} (\vec{r}) = (1/2) \left( \partial_{\alpha} u_{\beta} (\vec{r}) + \partial_{\beta} u_{\alpha} (\vec{r}) \right)$, where $u_{\alpha} (\vec{r})$ is the phonon displacement field at position $\vec{r}$. To include this full description, we also have to consider the 3 different phonon operators or equivalently the transverse and longitudinal modes of the phonon fields, and append $g_k$ with tensor indices $\alpha \beta$. Again, for the purpose of this paper, to get an estimate of the interaction strength between TLSs, we avoid this detail that does not change the power law of the interaction but simply introduces more tensorial structure, and show how one can attain the expected $1/r^3$ interaction via the polaron transformation method.    

We now consider the Hamiltonian of two TLSs (identified by indices $1$ and $2$ at positions $\vec{0}$ and $\vec{r}$ resp.) coupled to the phonon strain field as before,
\begin{align}
H &= \sum_{i=1,2} E^{z,o}_i \sigma^z_i  + \Delta^o_i \sigma^x_i + \sum_k g^k_1 (\sigma^z_1 + c^k_1) ( a_k + a^\dagger_{-k}) \nonumber \\
&+ \sum_k g^k_2 (\sigma^z_2 + c^k_2) (a_k + a^\dagger_{-k}) e^{i \vec{k}.\vec{r}} + \sum_k \om_k a^\dagger_k a_k.
\label{eq:unrem}
\end{align}
In what follows, we will include $e^{i\vec{k}.\vec{r}}$ in $g^k_2$. 

We now introduce the Polaron transformation, that allows for phonons to shift according to both TLS's states 
\be
U = exp \left( - \sum_k (a_k - a^\dagger_{-k}) \frac{1}{\om_k} ( \alpha^k_1 \sigma^z_1 + \alpha^k_2 \sigma^z_2 + \bk) \right) .
\label{eq:U}
\ee
 
For the Unitarity of $U$, we require $\alpha^k_i = (\alpha^{-k}_i)^{*}$, $\beta_k = \beta^*_{-k}$. This is supplemented with the Hermiticity of the Hamiltonian Eq. (\ref{eq:unrem}), that demands that $g^k_i = (g^{-k}_i)^{*}$. The transformed Hamiltonian under the polaron transformation is now 
\begin{align}
\tilde{H} &=  E_c + E^z_1 \sigma^z_1 + E^z_2 \sigma^z_2 + \Delta^o_1 \sigma^x_1 \avg{e^{-2 \alpha_1}} + \Delta^o_2 \sigma^x_2 \avg{e^{-2 \alpha_2}} \nonumber \\ 
&+ E_I \sigma^z_1 \sigma^z_2 + \sum_k ( V^k_1 \sigma^z_1 + V^k_2 \sigma^z_2 ) ( a_k + a^{\dagger}_{-k}) \nonumber \\
&+ \sum_k \om_k a^{\dagger}_k a_k + \sum_k ( a_k + a^{\dagger}_{-k} ) X_k  \nonumber \\
&+ \sum_{i=1,2} \Delta^o_i \sigma^+_i ( e^{-2 \alpha_i} - \avg{ e^{-2 \alpha_i}} ) +  \Delta^o_i \sigma^-_i ( e^{+2 \alpha_i} - \avg{ e^{-2 \alpha_i}} ), 
\label{eq:D3}
\end{align}

where $X_k = \beta_k - c^k_1 g^k_1 + c^k_2 g^k_2$. Again, before we can estimate the free energy, we need to eliminate the free linear term in the phonon fields, which requires shifting the phonon fields by $X_k$, as defined in Eq. (\ref{eq:D3}). As in Section \ref{sec:polaron}, this amounts to setting $\beta_k = c^k_1 g^k_1 + c^k_2 g^k_2$, thus eliminating $\beta_k$ from the variational procedure. The expressions for the terms in our Hamiltonian now reduce to (with $X_k = 0$)

\begin{align}
E_c &= \sum_k - \frac{\abs{ c^k_1 g^k_1 + c^k_2 g^k_2}^2}{\om_k} + (\alpha^k_1)^* \frac{(\alpha^k_1 - 2 g^k_1)}{\om_k}  \nonumber \\
&+  (\alpha^k_2)^* \frac{(\alpha^k_2 - 2 g^k_2)}{\om_k} ,\nonumber \\
E^z_i &= E^{z,o}_i - \sum_k 2 c^k_i \frac{\abs{g^k_i}^2}{\om_k} ,\nonumber \\
\Delta_i &= \Delta^o_i exp \left( - 2 \sum_k \frac{\abs{\alpha^k_i}^2}{\om^2_k} \coth{\beta \om_k / 2} \right) ,\nonumber \\
E_I &= \sum_k - \frac{ 2 (\alpha^k_1)^*}{\om_k} ( g^k_2 - \alpha^k_2 / 2) - \frac{ 2 (\alpha^k_2)^*}{\om_k} ( g^k_1 - \alpha^k_1 / 2) ,\nonumber \\
V^k_i &= g^k_i - \alpha^k_i.
\label{eq:res}
\end{align}

Thus, we find again that $E^z_i$'s are independent of the variational parameter $\alpha^k_i$, and that $\Delta_i$ is determined by the phonon displacements due to each TLS independently. However, if $E_I \ne 0$, then $\alpha^k_i$'s are coupled and the renormalization of $\Delta_i$ is indeed influenced by other TLSs.    

In principle, one must now minimize the free energy considering the spectrum of the Hamiltonian $H_{MF} = E_C + \sum_i E^z_i \sigma^z_i + \Delta_i \sigma^x_i + E_I \sigma^z_1 \sigma^z_2$, which is not a simple task. It is also important to remind oneself that we cannot interpret $E_I$ itself as the complete interaction strength between TLS 1 and TLS 2. The polaron transformation approach merely sets up an effective Hamiltonian in which the phonons have been shifted to their mean field positions. The remnant interaction with the phonon fields have zero expectation values, or are fluctuations which in principle can give an interaction between the TLSs. These in general involve considering the full dynamics of TLSs and phonons, and cannot be replaced by a simple effective Hamiltonian. Such a task is beyond the scope of this paper. 

The results of Black {\em et al}. however can be achieved by considering the regime in which spins do not have any dynamics, i.e, in the limit $\Delta \rightarrow 0$.  In such a scenario, the spins are essentially static and the system is classical. The only variables that depend on $\alpha^k_i$'s are now $E_I$ and $E_c$.  Regardless of the temperature of the system, minimization of free energy requires simply setting $\alpha^k_i = g^k_i$. This can be seen by noting that $\pd{E_c}{(\alpha^k_1)^*} = (g^k_1 - \alpha^k_1)$ and $\pd{E_I}{(\alpha^k_1)^*} = (g^k_2 - \alpha^k_2)$, that is, only proportional to $\alpha^k_i - g^k_i$. This simplification has the important consequence that the remnant interaction with phonons are also completely killed, $V^k_i = 0$ which allows us to understand $E_I$ as the complete interaction energy. In this scenario, we find -

\be
E_I = - \sum_k \frac{(g^k_1)^*g^k_2}{\om_k} + \frac{(g^k_2)^*g^k_1}{\om_k} .
\label{eq:eint2}
\ee

Now, for $g^k_1  = g_k$ and $g^k_2 = g_k e^{i \vec{k}.\vec{r}}$ with $g_k = \gamma_z k \sqrt{\hbar/ 2 \rho \om_k}$ as before, we find that the sum over $k$ in Eq. (\ref{eq:eint2}) is simply a fourier transform over something that is independent of any power of $k$, that is, $E_I = - \gamma_z^2/(\rho v^2) \sum_k e^{i \vec{k}. \vec{r}}$. This may seem peculiar in that, ordinarily this would imply a $\delta$ function interaction, but it must be noted that in 3 dimensions, the fourier transformation of any $T(\hat{r})/r^3 $ interaction is also independent of any power of $k$ but in general involves dependence on the direction of $k$. Suggestively, if we include anisotropies of the form $(i \hat{n}_1.\hat{k})$ and $(i \hat{n}_2.\hat{k})$ (with $\hat{n}_1$ and $\hat{n}_2$ defining dipole axes for TLSs 1 and 2 respectively) in $g^k_1$ and $g^k_2$ respectively, we arrive at the classical dipole-dipole interaction with the magnitude $\gamma_z^2/(4 \pi \rho v^2 r^3)$ and the appropriate tensor structure. 

For the purpose of this paper, it suffices to show that the polaron transformation can be used to yield the effective interaction between TLSs in the limit that $\Delta \rightarrow 0$, that is, upon neglecting the effect of TLS dynamics on phonon displacements. For the more complete result, the reader should look at the paper by Black and Halperin \cite{BH}, where upon obtaining the complete TLS-TLS interaction, they perform rms averaging over the angular structure of the interaction to arrive at an estimate of $C_{rms}$ in Eq. (\ref{eq:BH}). Finally, to arrive at the result in Eq. (\ref{eq:BH}) one simply rotates the basis to diagonalize the individual TLS Hamiltonians in Eq. (\ref{eq:D3}).

\bibliographystyle{apsrev4-1}
\bibliography{JJqubit2}

\end{document}